\documentstyle[numreferences]{spacekap}
\begin{opening}
\title{Match Probability Statistics and Gamma Ray Burst Recurrences in
        the BATSE Catalog}
\author {Sun Hong Rhie}
\institute {Institute of Geophysics and Planetary Physics, Lawrence
          Livermore National Laboratory, Livermore, CA 94550, U.S.A}
\author {David Bennett}
\institute {Center for Particle Astrophysics, University of California,
           Berkeley, CA 94720, U.S.A.}

\date {}
\end{opening}

\runningtitle{Match Statistics and GRB Recurrences in
        BATSE Catalog}

\begin{document}
\begin{abstract}
We develop {\it match probability statistics} to
 test the recurrences of gamma ray bursts in the BATSE catalog 1B and 2B.
We do not find a signal of repetitions at the match level of
$10^{-3}$\thinspace.
\end{abstract}

\vskip 0.5 truecm
\noindent
It is an interesting question whether classical gamma ray bursts
repeat, especially because of the potential implication on the distance
of the sources.  Gamma ray bursts were first discovered \cite{KSO} by Vela
satellites in 1967 the same year the first radio pulsar was
discovered.  Pulsars are  understood to be rotating neutron
stars in essence, but the origin of gamma-ray bursts is a complete mystery.
If they are from the (extended) Galactic halo, the isotropic
emission energy will be $\sim 10^{43}$ergs.  If they are from cosmological
distances, the isotropic emission energy will be $\sim 10^{51}$ergs.
For example, the bursts from cosmological distances can not be
repetitive because the enormous energy output would most likely be accompanied
by the consumption of the sources.

Currently, we do not have  dynamic clues toward a more meaningful
identification or rejection of repetitions, and
the controversy lies in whether there is statistical evidence
of repeating classical gamma ray bursts in the BATSE catalog.
In order to address the problem, we ask ourselves
the following questions.
 \begin{enumerate}
  \item{\it  Can we define a quantity which measures the probability
    for a given pair of GRB events to have come from the same position
    of the sky?}
  \item{\it  Can we find a statistic which can gauge the
       significance of matched pairs?}
  \item{\it  Is there a statistically significant excess of
       matched pairs in the BATSE catalog?}
 \end{enumerate}
As an answer, we
 construct the `match probability' of a given pair of events
based on measurement error distribution functions, build statistics
using `the match probability', and analyse the publically available
BATSE 1B and 2B data sets.

Our statistics is built based on the simple observation that the
repetitive bursts from the same source form an {\it error distribution}
in the position space. An obvious but important aspect of it is that
it is highly localised (around the true source position).
Gamma ray bursts in the BATSE catalog are known to be largely isotropic,
and the BATSE catalog can be considered
as a mixture of {\it error distributions}
and a {\it background  isotropic distribution}.
An isotropic distribution is characterized by the complete lack of
localization.  The drastic difference in localization behavior of
the two types of distributions is the underlying basis of our
burst pair match probability statistics.

Given a pair of bursts, the probability that they have come from the
same source  depends on the joint probability density of the error
functions of the events $f(\vec x_1)$ and $f(\vec x_2)$.
\begin{equation}
  \rho_{12} =  \int_{S^2} f_1(\vec x_1) f_2(\vec x_2)
            = \int_{S^2} f_1(\vec x - \vec x_{1c}) f_2(\vec x - \vec x_{2c})
              d^2\vec x  \ ,
\end{equation}
where $\vec x_{1c}$ and $\vec x_{2c}$ are the center positions.
If we assume gaussian error distributions
(in  $\Re^2$, which is good enough here),
\begin{equation}
  \rho_{12} = {1 \over 2\pi\sigma_{12}^2} e^{-r_{12}^2 /2\sigma_{12}^2}
\end{equation}
where $\sigma_{12}^2 = \sigma_1^2 + \sigma_2^2$.
We define the `match probability' of the pair by integrating $\rho_{12}$ from
the measured distance $r_{12} \equiv |\vec x_1 - \vec x_2|$ to infinity.
\begin{equation}
  P_{12} \ \equiv \ P (r_{12},  \sigma_{12})
         \ = \ {1\over \sigma_{12}^2}\  \int_{r_{12}}^\infty
            e^{-r_{12}^2 /2\sigma_{12}^2} \ r\, dr
         \ = \ e^{-r_{12}^2 /2\sigma_{12}^2}   \label{eq:pdef}
\end{equation}

Now, we look at the distribution of pairs in $P$-space.
For an {\it error distribution} in position space, the pair distribution
in $P$-space is homogeneous.
In other words, the number of pairs is a constant function of
$P$\thinspace.
Since $ P \in [0, 1]$,  the constant is the total number of pairs.
By normalizing by the total number of pairs, we define
the pair density distribution $g(P)$.
 \begin{equation}
  g_{\rm err} (P) = 1 \ .  \quad ; \qquad P \in [0, 1]
     \label{eq:gerr}
 \end{equation}
For the background isotropic distribution,
 \begin{equation}
  g_{\rm iso} (P) \propto {\sigma^2\over P}  \ .   \label{eq:giso}
 \end{equation}
(The number of pairs at a distance $r \propto r dr$ and
$r$ and $P$ are related by Equation~(\ref{eq:pdef}.)
It is not surprising that $g_{\rm iso} (P)$ is highly concentrated
toward small $P$ because {\it isotropic distribution}
has many more pairs at large separations in comparison to an
{\it error distribution} of the same number of bursts.

 $\sigma << 1$ for most of the bursts in the BATSE catalog,
and $g_{\rm err} (P) > g_{\rm iso} (P)$ unless  $P$ is very small.
Thus, we define the following statistical quantities.
 \begin{equation}
  \Delta Q_{\alpha}  \equiv  Q_\alpha(p)-<Q_\alpha(p)>_{\rm iso}
   \quad ; \alpha = 0, 1, ...
 \end{equation}
 \begin{equation}
  Q_{\alpha} (p) = \sum_{i<j} P_{ij}^\alpha \Theta(P_{ij}-p) \ , \\
 \end{equation}
where $\Theta (x) = 1 $ if $x > 0$ and $\Theta (x) = 0$ otherwise.
  The signal should be found in $\Delta Q_{\alpha}$, and
the background isotropic distributions for $<Q_\alpha(p)>_{\rm iso}$
are generated by randomizing the burst positions while keeping the
error functions of the data.  In the  limit of a large sample,
we can calculate $\Delta {Q_{\alpha}} (p)$.
 \begin{equation}
 \Delta {Q_{\alpha}}_{\rm th} (p) = {N_{\rm pairs} \over \alpha +1}
      \left( 1 - p^{\alpha+1} \right)
     - \sum_{\rm matched\ pairs} {\sigma_{ij}^2\over 2} \ .
 \end{equation}
Therefore,  we can calculate the expected signal to noise ratio ($S/N$)
by comparing $\Delta {Q_{\alpha}}_{\rm th} (p)$ to the RMS fluctuations
about $<Q_\alpha(p)>_{\rm iso}$.
We find that $Q_1(0)$ has the highest $S/N$.
For the fraction of the matched pairs (of bursts $\in$ error distribution)
of $10^{-3}$,  the $S/N$ of  $Q_1(0)$\ $ \approx 3.0$ for 485 bursts
(exclusive of the events without error estimations) of BATSE 2B.
We emphasize that our `match probability' test of the BATSE
catalogs for repetition is at the match level of $\approx 10^{-3}$.
For example,  if five bursts in the BATSE 1B
are from a repeater as suggested by Wang and Lingenfelder \cite{WL},
the signal can not be seen statistically because the required sensitivity is
$\approx 3 \times 10^{-4}$.
The analyses of $Q_1(0)$ statistics of the BATSE data are shown in Table 1.

\begin{table}[h]
\begin{center}
\caption{The number of matches and anti-matches measured
with the $Q_1(0)$ statistic with 1-$\sigma$ error bars determined from
the RMS deviation from the mean of 10,000 simulated BATSE catalogs. The
significance refers to the fraction of the 10,000 simulated catalogs with
larger values of $Q_1(0)$ than the real data.}

\begin{tabular} {r r r r r}
\hline
 Sample: & All & $\sigma_{\rm stat} < 9^\circ$
 &  $\sigma_{\rm stat} < 4^\circ$
 & Exposure Cut \\
\hline
$N_{\rm match}$    1B & $13 \pm 25$ & $27\pm 14$ & $ 1\pm 7$ & $3\pm 12$ \\
$N_{\rm antipode}$ 1B & $54 \pm 25$ & $24\pm 14$ & $17\pm 7$ & $9\pm 12$ \\
 match signif.     1B & 29.3 \% &  2.65\% & 43.36\% & 36.96\% \\
antipode signif.   1B &  2.42\% &  4.54\% &  1.32\% & 21.92\% \\ \hline
$N_{\rm match}$    2B & $-23\pm 45$ & $0 \pm 26$ & $-13\pm 13$ & $-9\pm 20$ \\
$N_{\rm antipode}$ 2B & $31 \pm 45$ & $39\pm 26$ & $29\pm 13$ & $1 \pm 20$ \\
 match signif.     2B & 68.31\% & 48.33\% & 82.39\% & 65.61\% \\
antipode signif.   2B & 23.34\% &  7.84\% &  1.90\% & 46.72\% \\
\hline
\end{tabular}
\end{center}
\end{table}

We conclude that there is no signal of recurrent bursts in the
BATSE data  at the match level of $10^{-3}$.
For more details, see \cite{BR}.

Now, where does our statistic lead to in regard to the
current controversy on the repetition of classical gamma ray bursts?
The outcome of devising a new statistic may not be of much value if it
merely provides a third statistic and another opinion on the issue.
It turns out that we can understand  both
two-point correlation  \cite{NP}
and nearest-neighbor statistics  \cite{QL}   from
`match probability' statistics because of the fuller position measurement
information contained in it and
moreover we can estimate the $S/N$ of the statistics.
Two point correlation function is directly related to $Q_0(p)$ and
the $S/N$ at the optimal $p_{\rm op}$ is somewhat worse than
the true optimal case of $Q_1(0)$.
One should keep in mind that the optimal value $p_{\rm op}$ can
be found only by embedding the two point correlation function in the
match probability statistics.
 In order to understand the nearest neighbour statistics (NNS),
we studied the cases of $60$ matched pairs.
We find that NNS has slightly better sensitivity than $Q_1(0)$ when each
source bursted only twice.
For multiplicities bigger than two, the $S/N$ of NNS degrades dramatically
and is far worse than that of the two point correlation statistics.
Therefore,  the conjecture by Quashnock and Lamb \cite{QL}
based on nearest neighbour correlations
that there are  recurrent bursts with high multiplicities  in the
BATSE catalog is not supported by more sensitive tests.
NNS also has a peculiar disadvantage that the sensitivity does not
improve as the size of the sample grows.

Finally,  we would like to  comment on the `signal'
Quashnock and Lamb \cite{QL} and  Narayan and Piran
\cite{NP}  have seen.  Is the `signal' identified in our analyses?
Yes:   In the sample $\sigma_{\rm stat} < 9^{\circ}$
( without exposure corrections),  the curve $Q_1(p)$ follows that of
${Q_1}_{\rm iso}$ and then deviates from it at around $p = 0.6$ to
follow the (theoretical) curve of signal for $10^{-3}$ matches.
Speaking in terms of $g(P)$, there is a fluctuation (depletion and excess)
in $g(P)$ at around $P = 0.6$ compared to that of $g_{\rm iso}$.
Is the `signal' consistent with the expectation from repeating bursts?
Clearly, it is not.
What is this `signal' a signal of?
We do not know.  It could be due to selection effects
\cite{NP,MAOZ}
 or many other reasons as yetuntested.
But it is also consistent with statistical fluctuations.
When we make exposure corrections using the trigger efficiencies provided
in the BATSE catalog,  the `signal' disappears.  We note that the events
excluded by this cut are mostly faint bursts.

\end{document}